\documentclass[useAMS,usegraphicx,usenatbib]{mn2e}
\usepackage{amssymb}

\newcommand{\lsun}{L$_\odot$}
\newcommand{\msun}{M$_\odot$}
\newcommand{\rsun}{R$_\odot$}

\def\lesssim{\mathrel{\hbox{\rlap{\hbox{\lower3pt\hbox{$\sim$}}}\hbox{\raise2pt\hbox{$<$}}}}}
\def\gtrsim{\mathrel{\hbox{\rlap{\hbox{\lower3pt\hbox{$\sim$}}}\hbox{\raise2pt\hbox{$>$}}}}}

\title[Hydro-Simulations of planet-star interactions]{Hydrodynamic
Simulations of the Interaction between Giant Stars and Planets}
\author[Staff et al.]
{Jan E. Staff$^1$\thanks{E-mail: jan.staff@mq.edu.au}, Orsola De Marco$^1$,
Peter Wood$^2$,
\newauthor Pablo Galaviz$^1$, and Jean-Claude Passy$^3$\\
$^1$Department of Physics and Astronomy, Macquarie University NSW
2109, Australia\\
$^2$Research School of Astronomy and Astrophysics, Australian National
University, Cotter Road, Weston Creek, ACT 2611, Australia\\
$^3$Argelander Institut f{\"u}r Astronomie, Bonn, Germany}

\begin{document}

\maketitle

\begin{abstract}

We present the results of hydrodynamic simulations of the interaction
between a $10$ Jupiter mass planet and a red or
asymptotic giant branch stars, both with a zero-age main sequence mass
of $3.5~{\rm M_\odot}$.  Dynamic in-spiral timescales are of the order of few years and a
few decades for the red and asymptotic giant branch stars, respectively.  The planets will
eventually be destroyed at a separation from the core of the giants smaller 
than the resolution of our simulations, either through evaporation
or tidal disruption.  As the planets in-spiral, the giant stars' envelopes
are somewhat puffed up.  Based on relatively long timescales and even
considering the fact that further in-spiral should take place before the
planets are destroyed, we predict that the merger would be difficult to
observe, with only a relatively small, slow brightening.  Very little mass
is unbound in the process.  These conclusions may change if the
planet's 
orbit enhances the star's main pulsation modes. Based on the angular momentum 
transfer, we also suspect that this star-planet interaction may be unable to 
lead to large scale outflows via the rotation-mediated dynamo effect of 
Nordhaus and Blackman.
Detectable pollution from the destroyed planets
would only result for the lightest, lowest metallicity stars.  We
furthermore find that in both simulations the planets move through the outer
stellar envelopes at Mach-3 to Mach-5,
reaching Mach-1 towards the end of the simulations.  The gravitational drag force
decreases and the in-spiral slows down at the sonic transition, as predicted
analytically.

\end{abstract}

\section{Introduction}

An increasing number of planets is being discovered at intermediate distances
from their host stars \citep{udry07}.   \citet{Villaver2009}, \citet{Mustill2012} and \citet{Nordhaus2013} among others calculated that expanding giants, whether red giant branch (RGB) stars or asymptotic giant branch (AGB) stars could engulf planets orbiting 2-4 times the maximum radius attained by the star. Such an interaction would likely result in the destruction of the planet producing an observational signature as well as long-lasting and observable evolutionary effects on stars.

Observational clues of star-planet interactions are in the form of putative planets discovered around post-main sequence stars, close enough that an interaction must have taken place when the star was in its giant phase in the recent past. Examples are the  
$1.25~{\rm M_J}$ planet (where ${\rm M_J}$ is the mass of Jupiter) 
orbiting 0.116 au from a horizontal branch star \citep{setiawan10}, two Earth-sized objects
orbiting a subdwarf B star at a separation of 0.0060 and 0.0076 au 
\citep{charpinet11}, or three earth-sized planets orbiting a subdwarf B pulsator \citep{Silvotti2014}. These
planets must have been engulfed in the envelope of the giant star that
became the subdwarf star today. \citet{charpinet11} and \citet{passy12planet} showed how
planets may have been much more massive initially and lost much of their
mass in the CE phase. However, they could not determine how the core of the planets survived the interaction instead of plunging into the core of the giant. 

A second type of planet around post-giant stars consists of one or more
planets detected at au-distance from post-CE binaries, rather than single
stars.  Some of those finds have been debated because the planets would not be in stable orbits \citep{Horner2013} or the data could be as easily 
explained with alternative, non-planet scenarios.  However, other data is
more convincingly, though not conclusively, explained by the presence of
planetary systems \citep[e.g., NN~Serpentis;][]{Parsons2014}.  These
planets, contrary to the planets at sub-au orbital separations from {\it
single} post-giant stars, are unlikely to have been involved in the CE that
created today's close binary, but they must have been impacted by the
ejection of the giant's envelope.  It has been speculated that they may have
formed in the aftermath of the CE binary interaction \citep{Beuermann2011},
similarly to how the planets around pulsar PSR1257+12 \citep{Wolszczan1992}
were formed after the supernova explosion.

Theoretically \citet{soker98} suggested that star-planet interactions could 
generate blue horizontal branch stars by enhancing the RGB mass loss rate and 
decreasing the envelope mass of the red giant star.  \citet{carlberg09} 
calculated instead the extent to which interactions (tidal interactions 
and/or mergers) between giants and planets would spin up giants. 
\citet{nelemans98} found
that companions with masses more than $20-25~{\rm M_J}$ could survive a common
envelope with a $1~{\rm M_\odot}$ red giant and expel the envelope. 
On the other hand, using the formalism from \citet{nelemans98},
\citet{villaver07} found that companions less massive than $120~{\rm
M_J}=0.11~{\rm M_\odot}$ would evaporate inside the envelope of a $5~{\rm
M_\odot}$ AGB star.  Both these studies rely on the uncertain and highly
debated efficiency of the common envelope ejection formalism
\citep{DeMarco2011}.  

 \citet{nordhaus06} investigated analytically $3~{\rm M_\odot}$ RGB and AGB stars
interacting with a low mass companion (planet, brown dwarf, or low mass
main-sequence star). They found that envelope ejection in the RGB case is
unlikely, but possible for AGB stars from low-mass main sequence stars. 
While planets may have too low a mass to
eject the AGB star's envelope, they found that a planet may induce
differential rotation mediated dynamo that can eject material. Furthermore,
the planet may tidally disrupt, creating a disc inside the envelope that can
lead to a disc driven outflow.
\citet{metzger12} investigated mergers between hot
Jupiters and their host main sequence stars and predicted that prior to
merger, as the planet penetrates the star's atmosphere, an EUV/X-ray
transient is produced in the hot wake following the planet.  The merger
would also drive an outflow and hydrogen recombination in the outflow would
cause an optical transient.  They argued that the galactic rate of mergers
between hot Jupiters and their host stars should be $0.1-1~{\rm yr^{-1}}$
and that should be similar to the rate observed for planets and giant stars.

Despite past efforts, many questions still remain. What is the fate of the planet in the CE
interaction?  Does it survive, or is it destroyed by ablation or tidal
disruption?  Whatever the fate of the planet, will the interaction lead to
an alteration of the star and its subsequent evolution, such as spin-up,
mass loss, or a change in the surface abundances?

Presumably, once a planet is tidally brought closer to an expanding giant star, the star would fill its Roche lobe and transfer mass to the planet. Since the planet is much less massive than the star, it is likely that the planet would be engulfed by the giant's extended envelope and have a common envelope (CE) interaction  \citep{Ivanova2013}. CE interactions are thought to happen also between giants and stellar mass companions \citep{Paczynski1976} and give rise to compact evolved binaries.  CE simulations using a variety of techniques
\citep[e.g.][]{sandquist98,ricker12,passy12,nandez14,staff15} have a range of uncertainties and shortcomings  \citep[e.g.,][]{Nandez2015}, but can be used as starting points to determine the nature of star-planet interactions.
By running hydrodynamic simulations of the CE interaction
between a $10~{\rm M_J}$ planet and an RGB or an AGB star, we start addressing numerically aspects of the interaction such as the timescale of the interaction, the final separation or the extent to which the stellar envelope is spun-up.  

We also exploit the lightness of the planetary companion relative to the
stellar envelope to carry out a study of gravitational drag experienced by a
body in a common envelope simulation \citep{Ricker2008}.  This is much more
difficult when the companion is more massive because the gas is stirred
considerably and it is difficult to extract some of the quantities needed to
carry out the calculation.

We describe the numerical method that we use in 
section~\ref{methodssection}. Then in
section~\ref{resultssection} we present our results including an appraisal of how numerical considerations impact our conclusions. In Section~\ref{dragforcesubsection} we assess the drag forces acting on the planet and we exploit the relative composure of the envelope gas to compare these forces to their analytically-derived equivalent. We finally discuss our results in Section~\ref{summarysection}.

\section{Method}
\label{methodssection}

\begin{figure*}
\includegraphics[width=0.9\textwidth]{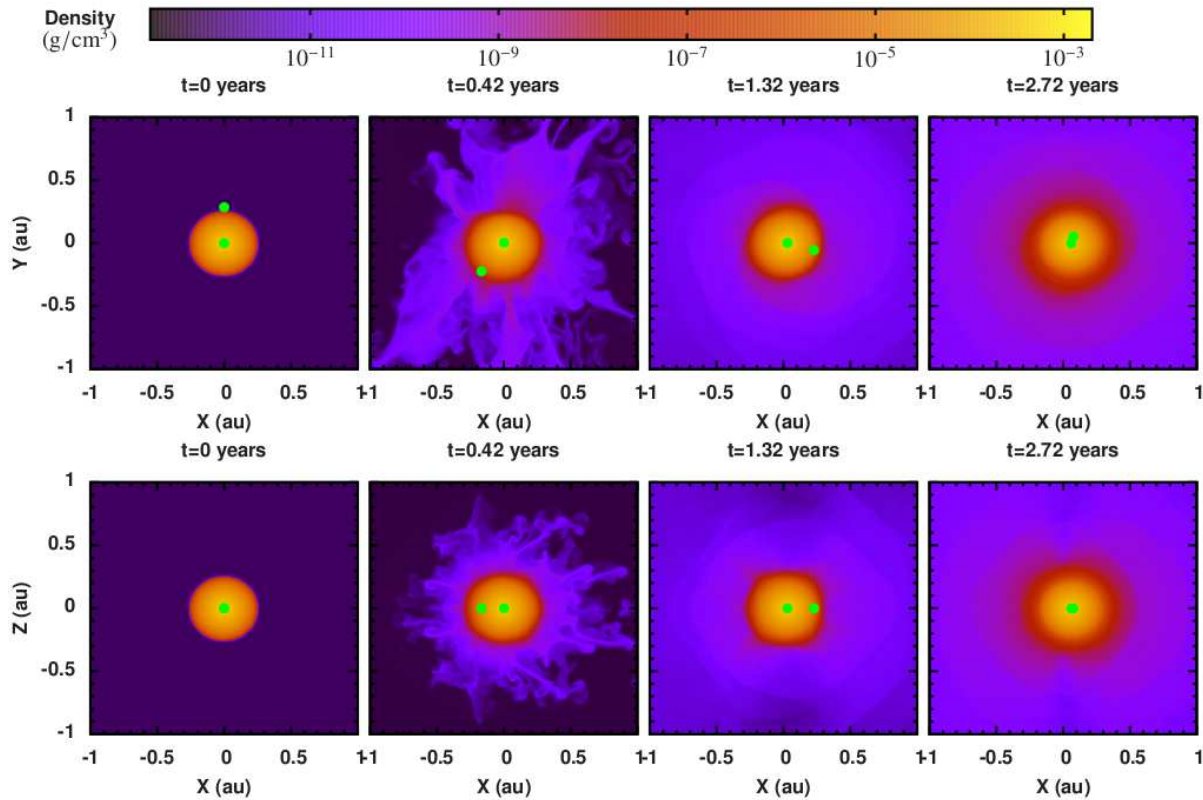}
\caption{Density slices taken through the middle of the grid on the
perpendicular (upper panels) and orbital (lower panels) planes, at 
four different times, for the low resolution simulation with the RGB star. The leftmost 
column shows the initial setup. The core of the RGB star and the planet
are indicated by green dots.}
\label{rgbrhopanels}
\end{figure*}
\begin{figure*}
\includegraphics[width=0.9\textwidth]{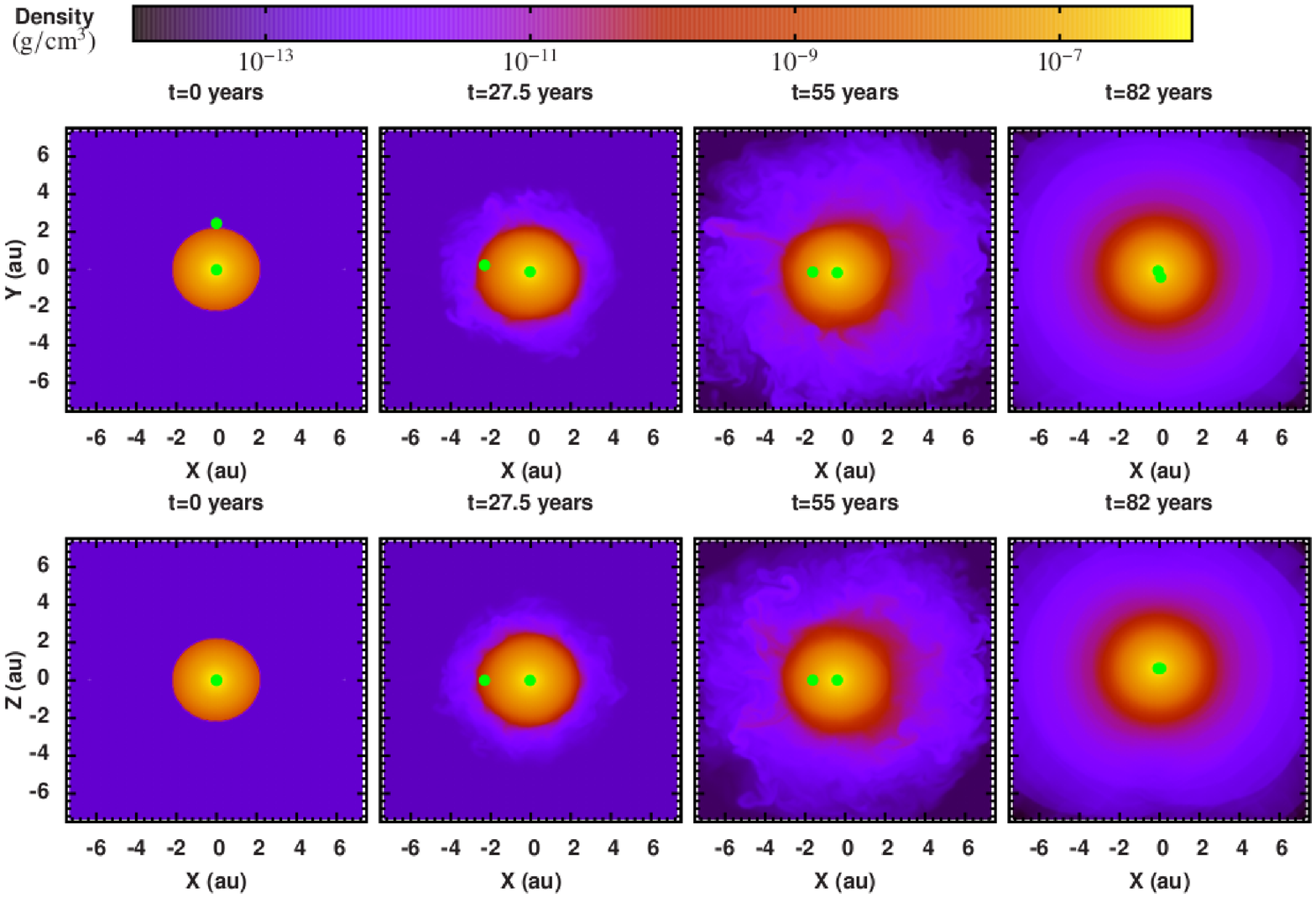}
\caption{Density slices taken through the middle of the grid on the
perpendicular (upper panel) and orbital (lower panel) planes,
at four different times, for the low resolution simulation with
an AGB star and a planet. The leftmost column of panels is the initial setup
in the three different cuts, then at $t=27.5~{\rm years}$, $t=55~{\rm
years}$, and $t=82~{\rm years}$. The core of the RGB star and the planet
are indicated with green dots.}
\label{agbrhopanels}
\end{figure*}

We used a modified version of the grid-based hydrodynamics code \textsc{enzo}
\citep{oshea04,passy12,bryan14} to run the hydrodynamics simulations.  The
calculations were performed on a $256^3$ grid in the adiabatic approximation
with outflow boundary conditions. In addition, we also performed the same
simulations on a grid with $512^3$ resolution, to test if the resolution 
affects the results.

The structure of the giant stars were calculated using the stellar evolution
code Modules for Experiments in Stellar Astrophysics 
\citep[\textsc{mesa} ;][]{paxton11,paxton13}.  We used two stellar structures,
evolved from the same 3.5-\msun, zero-age main sequence, solar metallicity
model.  The first stellar structure was that of the model 
283 million years after joining the zero-age man sequence.  
At this time the star had reached the RGB, having
lost 0.01~\msun.  It had a radius of 55~\rsun\ (approximately the maximum
radius that this type of star reaches on the RGB), a luminosity of
680~\lsun, and an effective temperature of 3960~K.

The second stellar structure was taken 330 million years after the zero-age
main sequence.  At this time the star had reached the thermally-pulsating
AGB, had a mass of 3.05~\msun\, a radius of 470~\rsun, a luminosity of
$1.4\times10^4~{\rm L_\odot}$ and an effective temperature of 2920~K (the
structure was taken between two thermal pulses).  This is the same stellar
structure used for the simulations of \citet{staff15}.  
Stars more massive than $\sim2~{\rm M_\odot}$
grow a lot larger on the AGB than on the RGB, providing for
an opportunity for planets that had not interacted on the
RGB to do so during the AGB. This is the reason why we
chose to use a star more massive than $2~{\rm M_\odot}$.
Stars less massive than approximately $2~{\rm M_\odot}$ have similar
maximum RGB and AGB radii and this means that they have
most of their interactions on the RGB.
For the more luminous
low mass ($M < 2~{\rm M_\odot}$) RGB stars, which can attain radii close
to $200~{\rm R_\odot}$, the nature of the star-planet interaction will be
intermediate between the RGB and AGB cases considered here.

We mapped the 1-D \textsc{mesa} model into the \textsc{enzo} computational
domain.  \textsc{mesa} models have much higher resolution compared to the
linear resolution of the 3-D Cartesian \textsc{enzo} grid that we use.  The
size of the simulation box used for the simulation with the smaller RGB star
was $3\times10^{13}~{\rm cm}$ (2 au), such that each cell in $256^3$
cell domain had a size of
$1.2\times10^{11}~{\rm cm}$ (1.7~\rsun).  In the simulation with the larger
AGB star, the simulation box was $2.2\times10^{14}~{\rm cm}$ (15 au), and
each cell in $256^3$ resolution had a size of $8.6\times10^{11}~{\rm cm}$ 
(12~\rsun). The cell size was half these values for the $512^3$ resolution simulation.  The
cores of the giant stars, where much mass is concentrated, as well as the planet
companion cannot be resolved.  Instead, they are
approximated by point masses, with a smoothed gravitational potential as
discussed in \citet{passy12} and by \citet{staff15}.  We used a smoothing
length of 3 cells, instead of 1.5, which, as was discussed in
\citet{staff15}, results in better energy conservation.

\textsc{mesa} takes microphysics into account, while we use an ideal
gas equation of state with an adiabatic pressure-density relation 
($\gamma=5/3$) in \textsc{enzo}.  Because of this
and the addition of a point mass with a smoothed gravitational potential,
the star is not in perfect hydrostatic equilibrium in \textsc{enzo}. 
Following the method described by \citet{passy12}, we force the starting
model to hydrostatic equilibrium by dampening the velocities by a factor of
3 for each time step after mapping the stellar structures into the
computational domain.  We then check the stability by running the simulations
without damping the velocities for 4 dynamical times (the dynamical times are
0.07 years  and 1.8
years for the RGB and AGB stars, respectively).  The simulation volume not
occupied by stellar gas is filled with a hot medium, which has a density
four orders of magnitude lower than the giant star's least dense point and a
high temperature so as to balance the pressure at the surface of the giant
star.  
Despite this, the outer layers of the star tend
to diffuse out somewhat \citep[see][]{staff15}.
The 3D star constructed in this way tends to be slightly larger than
it was initially.  For both models the post-stabilization radius was
$\approx$5 per cent larger
(2.5~\rsun\ and 23~\rsun~larger for the RGB and AGB models, respectively, at a density
one order of magnitude less than the initial lowest density in the star).  

Once the giant star is stabilized, we insert a planet with a mass of
$10~{\rm M_J}$ at $1.1$ times the radius of the \textsc{mesa} model ($R_{\rm
star}$), in a circular orbit.  In both simulations this initial
configuration results in the giant stars massively overflowing their Roche
lobe radii.  This is the case with many CE simulations
\citep[e.g.,][]{sandquist98,passy12} and may have some effect on the CE
outcome (Iaconi et al., 2016, in preparation).  However, in the case of planetary companions, it is likely that
the effect of starting close to the surface is minimal:  companions as far
as 2-3 stellar radii are likely to be captured \citep{Villaver2009,Mustill2012}, but the angular momentum of the orbit  transferred to the
primary would confer to it only a relatively minor surface velocity of $1.1-1.3~{\rm
km~s^{-1}}$ for the AGB star and $3.2-3.9~{\rm km~s^{-1}}$ for the RGB star 
(this range was found assuming that all the
orbital angular momentum of the planet at a distance of 2-3 stellar radii
is transferred to the envelope of the giant, and that this envelope rotates
rigidly), not too different from our non-rotating initial models.

\section{Results: the in-spiral}
\label{resultssection}

\begin{figure}
\includegraphics[width=0.45\textwidth]{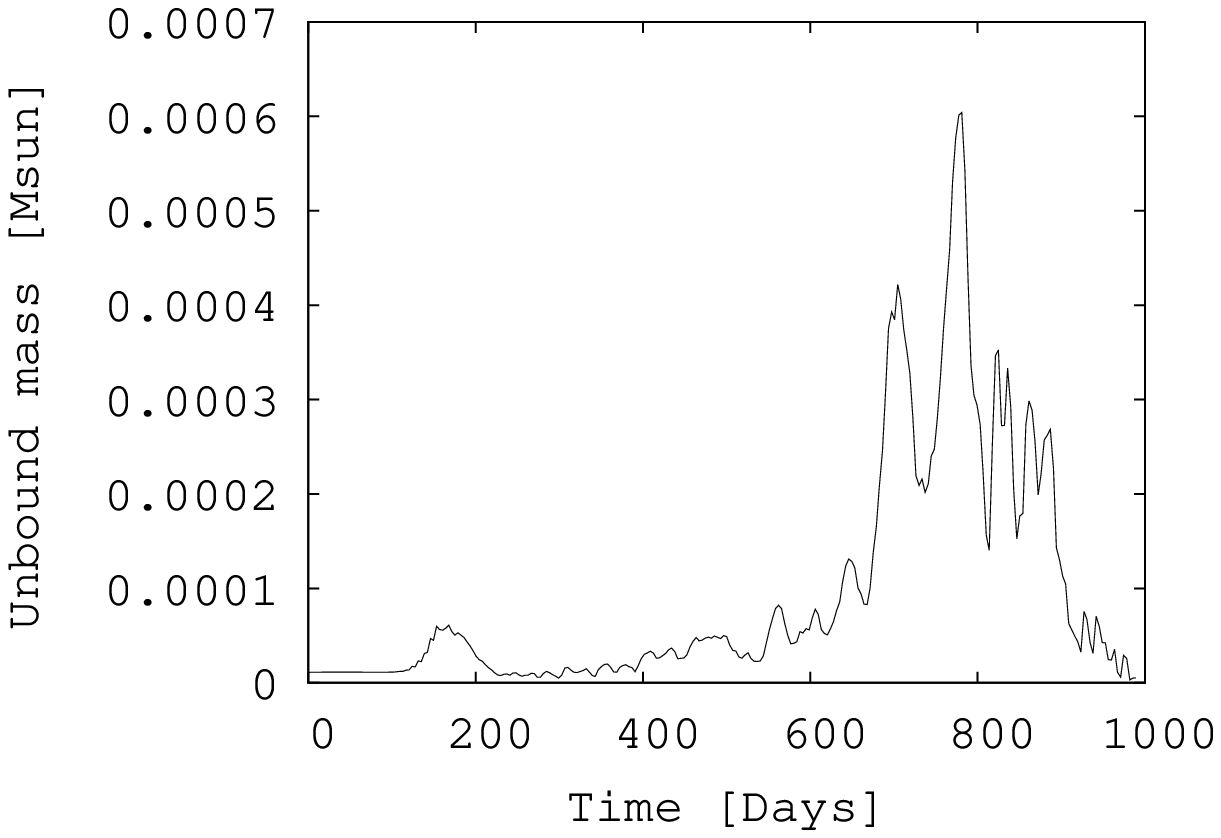}
\includegraphics[width=0.45\textwidth]{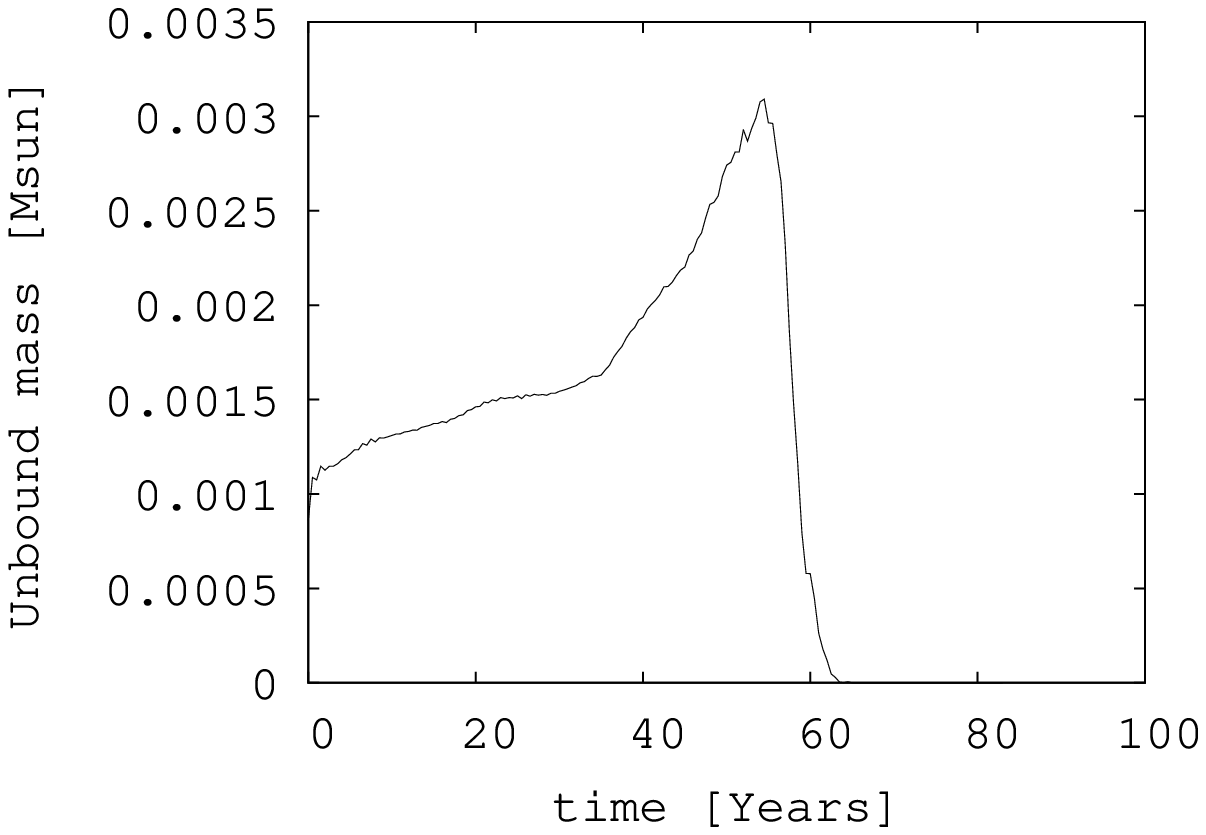}
\caption{The unbound mass in the computational domain as a function 
of time for the RGB star (upper
panel), and the AGB star (lower panel) orbited by a
$10~{\rm M_J}$ planet (in both cases the figures are from the low 
resolution simulations).}
\label{rgbmunb}
\end{figure}

With our starting conditions ($a_0 = 1.1 R_{\rm star} = 61~{\rm
R_\odot}=0.28~{\rm au}$, and an orbital period of 26 days for the RGB star
and $a_0 =1.1~R_{\rm star}=520~{\rm R_\odot}=2.4~{\rm au}$ and an initial
period of 1.9~years for the AGB star), the planet is rapidly engulfed by
stellar envelope gas.  We show the evolution of the density 
for the RGB star in Fig.~\ref{rgbrhopanels} and for the AGB star
in Fig.~\ref{agbrhopanels}.  As the planet in-spirals, the giant star's
envelope expands.  This puffed-up envelope has typical densities of
$\sim10^{-10}~{\rm g~cm^{-3}}$ in the RGB simulation and $\sim10^{-12}~{\rm
g~cm^{-3}}$ in the AGB simulation.  As the stellar envelope is puffed-up due
to the interaction, the photosphere is likely to be located
near the edge of this expanding gas. Due to the high temperature
ambient medium, this low-density puffed up gas may be artificially heated.
Therefore we cannot accurately determine the temperature of the
photosphere, nor how fast it would cool off radiatively and therefore
recede. Especially in the RGB simulation, where the interaction is
reasonably quick, it seems likely that a significant increase in the
photospheric radius could be achieved.

Some low density gas is lost from the domain, $\lesssim0.01~{\rm M_\odot}$
in both simulations.  Of the mass lost from the simulation box, $\lesssim10$
per cent ($\approx10^{-3}~{\rm M_\odot}\approx1~{\rm M_J}$) is unbound in
the RGB simulation, and $\lesssim30$ per cent ($\approx3\times 10^{-3}~{\rm
M_\odot} \approx3~{\rm M_J}$) is unbound in the AGB simulation (see
Fig.~\ref{rgbmunb}).  We note that initially, a larger amount of the ambient
medium is unbound in the AGB simulation compared with the RGB simulation,
which in part explains why the AGB simulation unbinds more mass. 
Pre-empting our discussion on energy conservation in Section~3.1, we note
that there is considerable uncertainty on the mass unbinding.  The change in
the planet's orbital energy as it spirals through the RGB star layers, that
can lead to unbinding of envelope, is $\sim3\times10^{45}~{\rm erg}$, an
order of magnitude smaller than the artificial growth in the total energy in
the box for that simulation.  This artificial growth in energy may therefore
be the main driver for the meagre mass unbinding observed, making our
estimate for the RGB case an upper limit. 

\begin{figure*}
\includegraphics[width=0.44\textwidth]{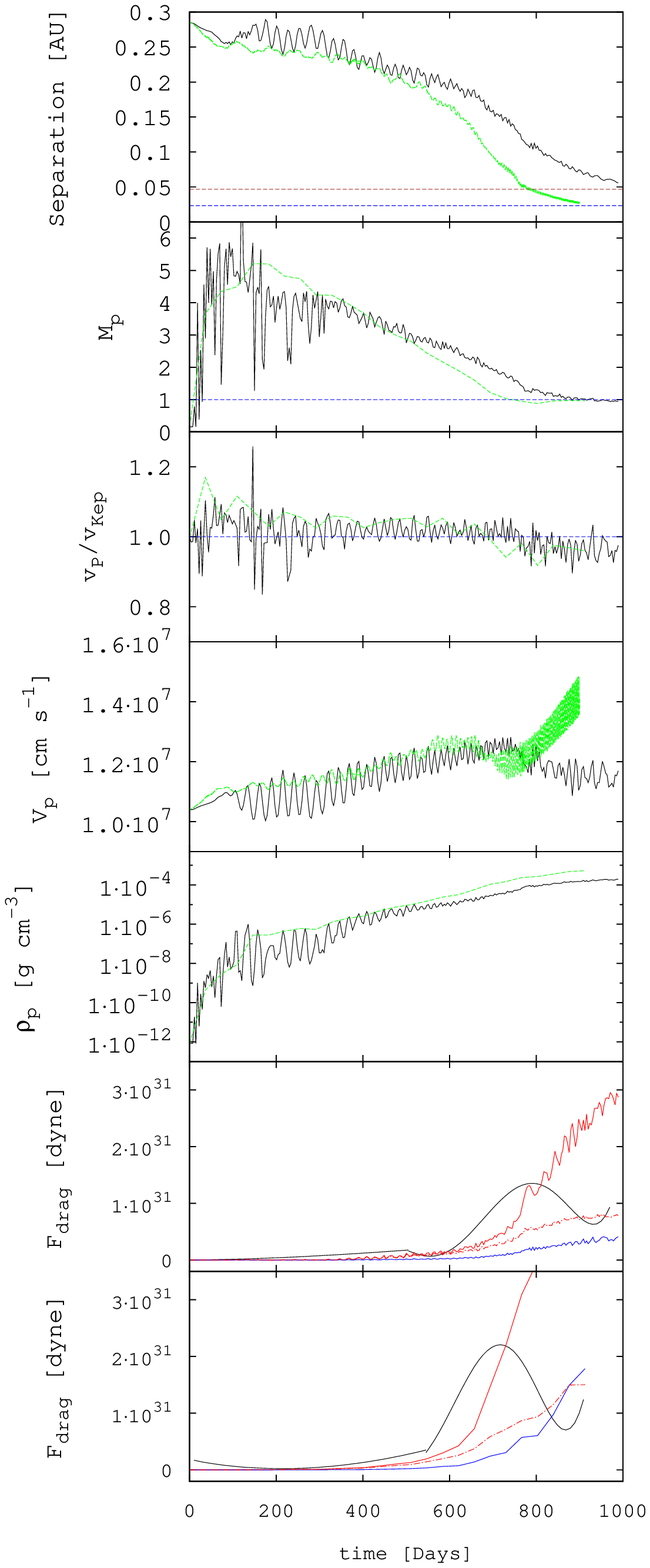}
\hspace{1cm}
\includegraphics[width=0.44\textwidth]{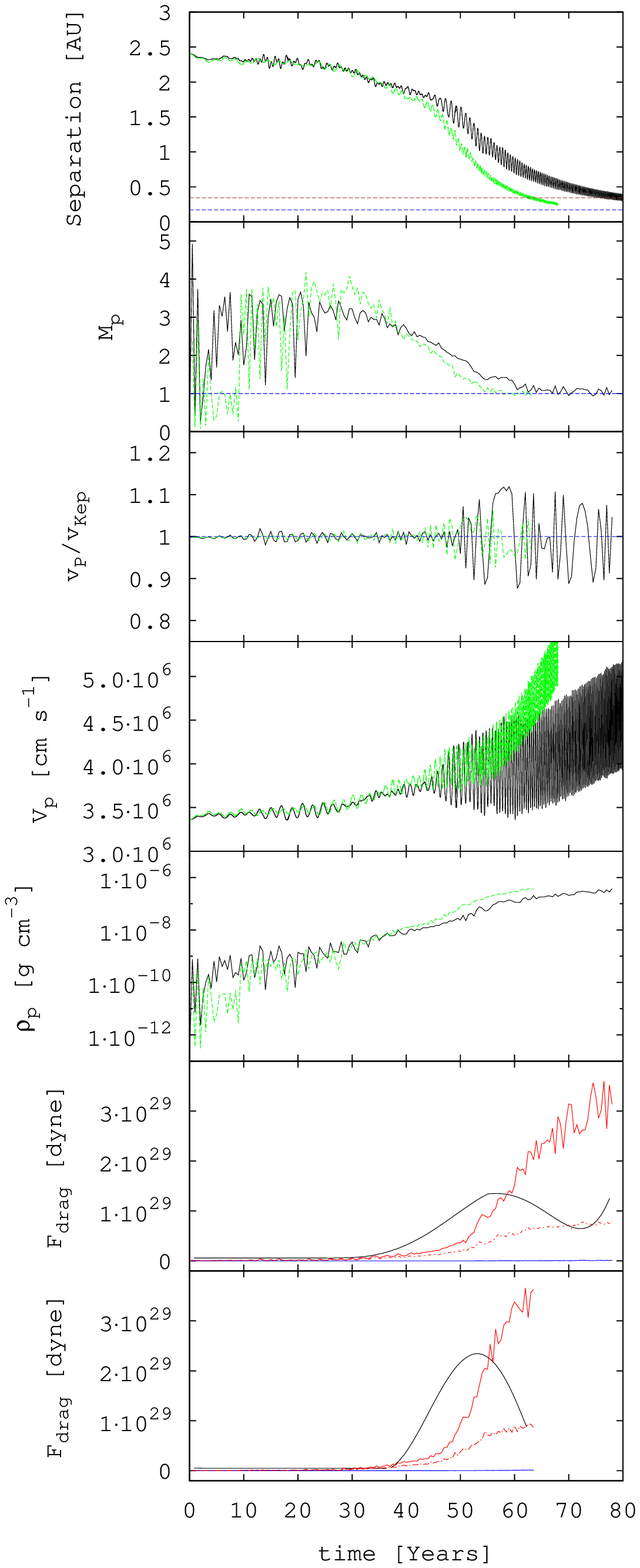}
\vspace{1cm}
\caption{Results from the RGB simulation (left column) and AGB simulation
(right column). Black curves are results from the low resolution
simulations. Green curves are from the high resolution simulations. 
{\it Upper panel:} separation
between the core and the planet as a function of time. The blue dashed line
represents two smoothing lengths in the higher resolution simulation, while
the brown dashed line represents two smoothing lengths in the lower
resolution simulation. {\it Second panel:}
the Mach number of the planet as a function of time. The dashed blue line
indicates $M_{\rm p}=1$.{\it Third panel:} the
planet's velocity relative to the Keplerian velocity. The dashed blue line
indicates $v_{\rm p}/v_{\rm Kep}=1$. {\it Fourth panel:}
the planet's velocity relative to the grid. {\it Fifth panel:} the density of the stellar
envelope around the planet. {\it Sixth panel:} the drag force
calculated in the low resolution simulations (black curve), compared with the 
gravitational drag force calculated from the analytical
expression including pressure effects (Eq.~\ref{gravdrageq}; dashed
red curve), and excluding pressure effects (solid red curve), as well as the
hydrodynamic drag force (blue curve) calculated from the analytical expression
(Eq.~\ref{hydrodrageq}). {\it Seventh (bottom) panel:} the same as the sixth
panel, but for the high resolution simulations.}
\label{multifigure}
\end{figure*}

\begin{figure}
\includegraphics[width=0.45\textwidth]{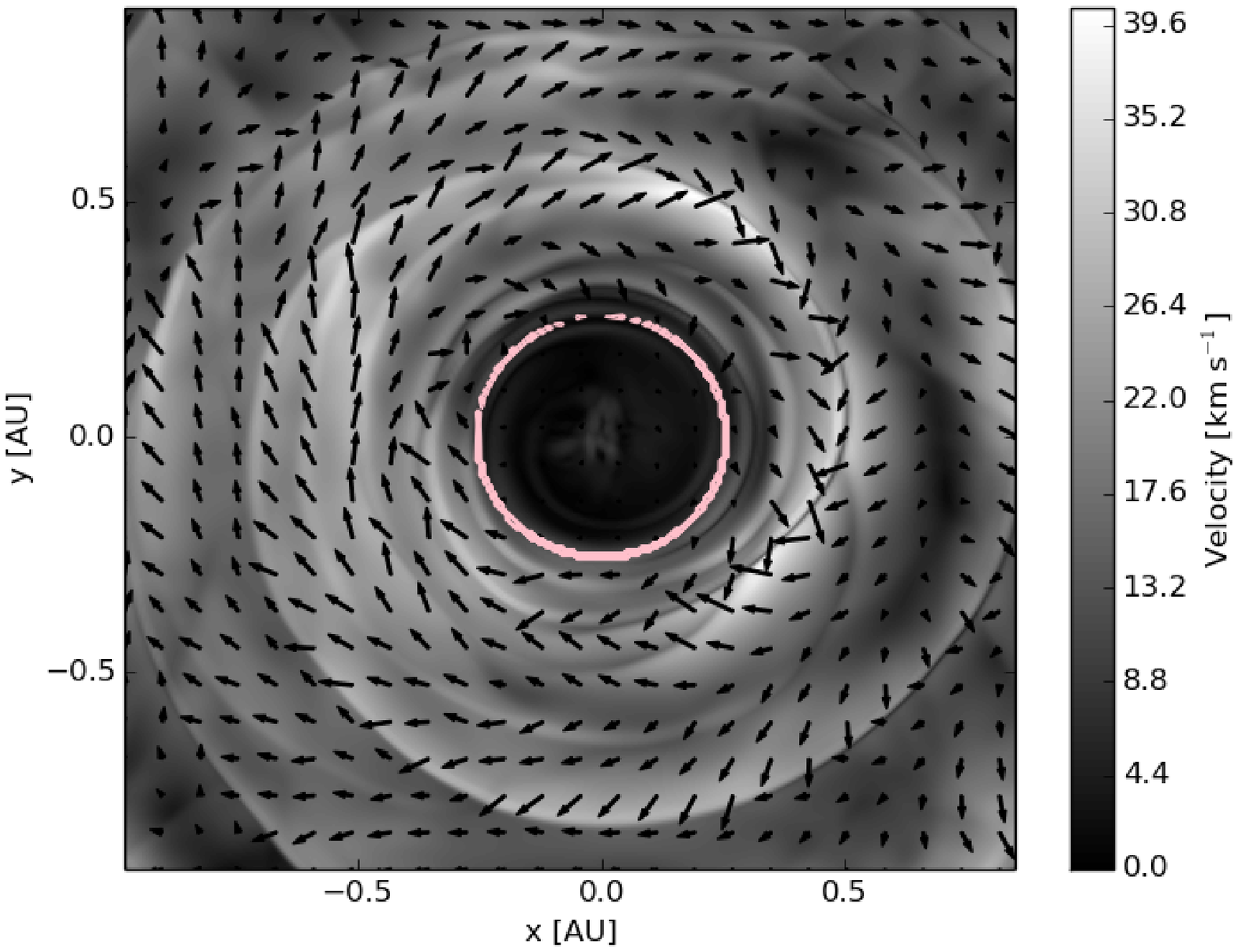}
\includegraphics[width=0.45\textwidth]{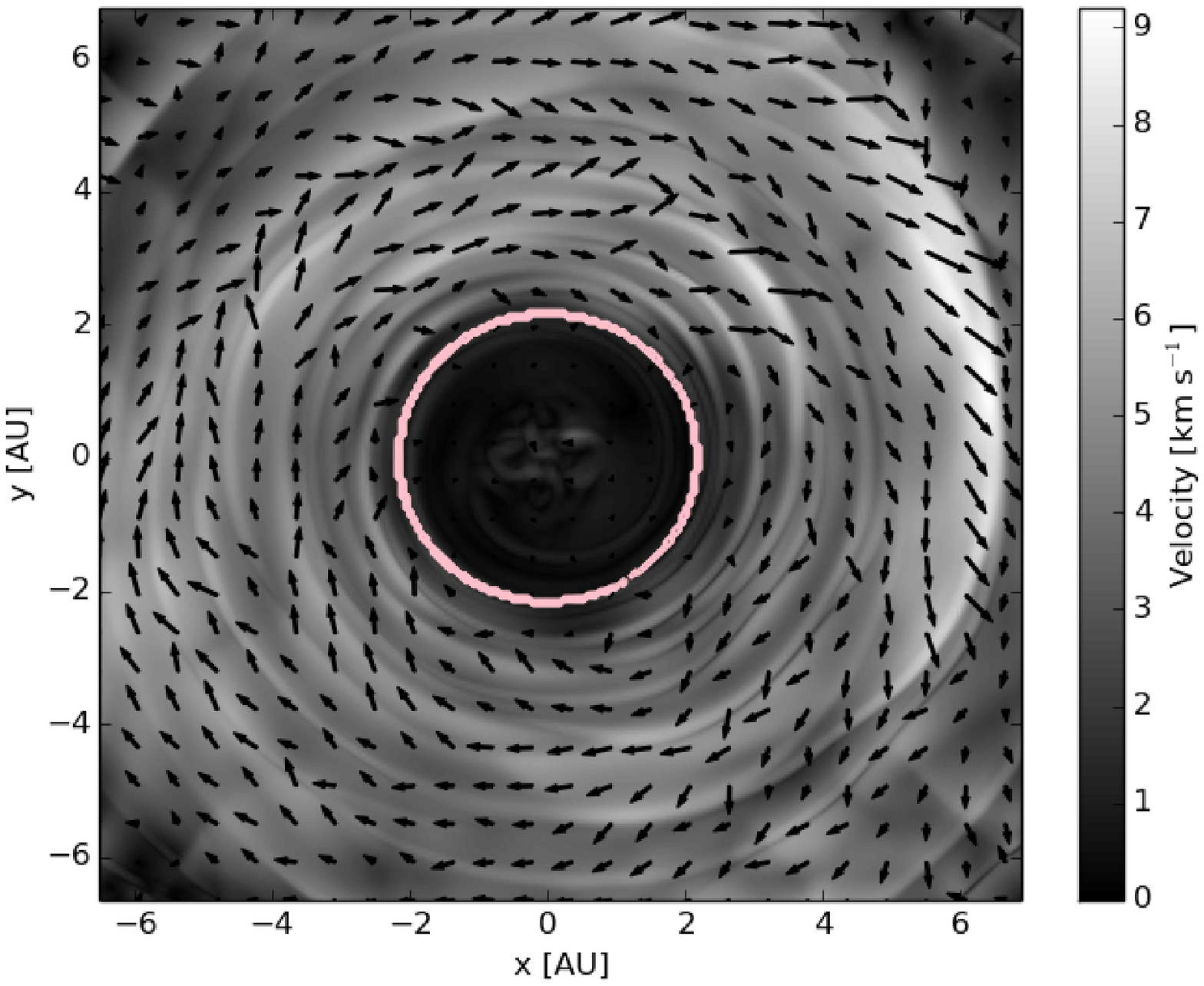}
\caption{The velocity of the gas in the equatorial plane at the end of the
low resolution simulations. {\it Top panel:} RGB star, {\it bottom panel:} AGB star. The
plotted velocity is $\sqrt{v_x^2+v_y^2}$, with the arrows showing the 
direction. The pink circle indicates the size of the giant star prior to
the interaction. It is only the puffed up, low density matter at larger radii 
that gains
significant rotational velocity, while the high density interior of the
stars have very low rotational velocity.}
\label{rotvelfig}
\end{figure}

Shown in Fig.~\ref{multifigure} are a series of quantities, which we describe
and compare in detail below, for both low and high resolution, RGB and 
the AGB simulations.  In the top panel we show the separation
between the planet and the core of the giant star as a function of time,
then we show the planet's Mach number ($M_{\rm p}$) as it moves through the 
stellar envelope; third is its velocity with respect to the Keplerian value
($v_{\rm Kep}$), 
then its velocity with respect to the grid, followed by the envelope
density surrounding the planet, and, finally, the gravitational drag force
acting on the planet in the low resolution simulation (sixth panel) and the
high resolution simulation (seventh or bottom panel). The last two panel rows will be exhaustively discussed in Section~\ref{dragforcesubsection}.
The sound speed used
to compute the Mach number is just the sound speed in the cell in which the
particle is located.  Likewise, the density and velocity of the gas surrounding the particle are the values of the cell where the particle is located.

The overall behaviour of the
separation is a gradual decrease over 2-3 years in the RGB
simulation, and over approximately 60-80 years in the AGB simulation (faster
for the higher resolution simulations), after which we
cannot follow the evolution because the separation approaches 0.05 and 0.4~au for the RGB and AGB simulations, respectively, which is
close to two smoothing lengths (one smoothing length in the lower resolution
simulations is
$3.5\times10^{11}~{\rm cm}=0.023~{\rm au}$ for the RGB star and
$2.6\times10^{12}~{\rm cm}=0.17~{\rm au}$ for the AGB star), at which point 
the smoothing of
the potential may begin to impact the results (see also
Section~\ref{numconssection}). 

The oscillatory behavior seen in the separation plot for the  low
resolution RGB star in
Fig.~\ref{multifigure} has a period of $\approx25$ days (similar to the
planet's initial orbital period) and is due to the development of an
eccentricity, typically observed during the fast in-spiral phase of CE
simulations and ascribed to the non symmetric distribution of gas \citep[see
e.g.][]{passy12}.  Between $\sim100$ and $\sim300$ days in the RGB
simulation, the orbital separation and the planet's velocity remain
approximately constant.  Following this, the planet speeds up as the
separation decays. In the higher resolution RGB simulation, the snapshots
from the hydrodynamics simulation
were produced less frequently, with a frequency of $0.1$ years (which is
larger than the oscillatory period), and this
oscillatory behaviour is therefore partly hidden in Fig.~\ref{multifigure}.
In the AGB simulation, the separation also remains
approximately constant for the first $\sim 30$ years and the orbit develops
a lower eccentricity than for the RGB case.  After this, the separation
decays, and between 50 and 60 years there is a rapid decrease in the
separation.
Although
we observe a period of faster in-spiral between 700 and 800 days in the RGB
simulation, this is
not as prominent as in past CE simulations or in the CE between the planet
and the AGB star. It is however similar to the behavior
of a 0.01~\msun\ companion plunging into the 0.88~\msun, 85~\rsun~RGB star
(De Marco et al.  2012).

During the interaction the outer layers of the puffed up envelope gain rotation. At
densities lower than the initial photospheric density ($\rho<8\times10^{-9}~{\rm g~cm^{-3}}$ for the RGB star and $\rho<1\times10^{-9}~{\rm g~cm^{-3}}$ for the AGB star), 
rotational velocities of
$\gtrsim20~{\rm km~s^{-1}}$ are found in the RGB simulation, and
$\gtrsim5~{\rm km~s^{-1}}$ are found in the AGB simulation 
(see Fig.~\ref{rotvelfig}). But since the planet has a low angular momentum
due to its low mass, the planet is unable to 
noticeably spin up the higher density, more massive, layers of the giant stars. At
higher densities than the photospheric density of the initial model the star is
therefore not rotating. 

The velocity of this puffed-up envelope is, however, small compared to the
planet's velocity around the giant star, and 
the planet's orbital velocity with respect to the grid is therefore similar 
to the velocity relative to the surrounding gas in both simulations. 
The planet's velocity early in the simulation is seen to oscillate between
$100$ and $120~{\rm km~s^{-1}}$ in the RGB case,
while in the AGB simulation the orbital velocity of the planet varies less
around a value of
 $\approx 35~{\rm km~s^{-1}}$.  During the fast in-spiral, at approximately
600 days in the RGB simulation, the velocity continues to oscillate and
increases up to $\approx130~{\rm km~s^{-1}}$, while in the AGB
simulation the velocity is seen to oscillate more towards the end of the
simulation, reaching a maximum of $\approx50~{\rm km~s^{-1}}$.  As the
planet in-spirals, its velocity is found to remain approximately Keplerian
throughout both simulations.

Once the planet becomes submerged in the stellar envelope, the Mach number
jumps to 4 or 5 in both simulations (see Fig.~\ref{multifigure}).  
During the rapid in-spiral phase, the
Mach number decreases as the sound speed grows deeper inside the giant stars.  At the end of
both simulations, the planet's velocity is
approximately the same as the sound speed. 
In Sec.~\ref{dragforcesubsection} we will discuss the drag force in relation to the Mach number of the particles.

\subsection{Numerical considerations}
\label{numconssection}

\begin{figure}
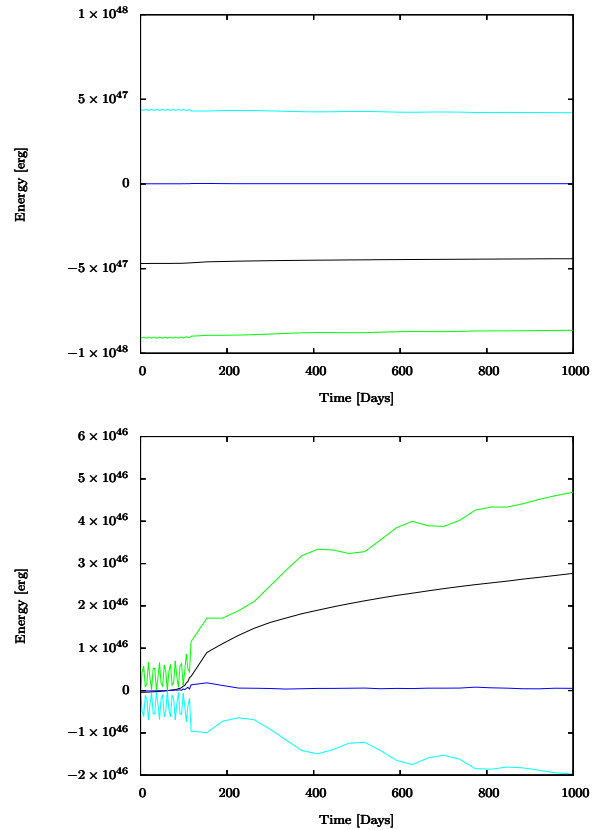

\resizebox{0.45\textwidth}{!}{\input{energies3.5rgb510mj1.1.tex}}
\resizebox{0.45\textwidth}{!}{\input{energies3.5rgb510mj1.1zoom.tex}}
\caption{{\it Top panel:} Energy components on the grid in the lower 
resolution RGB 
simulation as a function of time ({\it cyan curve}:
thermal energy, {\it blue curve}: kinetic energy, {\it black curve}: total 
energy, and 
{\it green curve}: gravitational potential energy).
{\it Bottom panel:} The energy components 
shifted so that the curves start at 0, to illustrate the differences over
time.}
\label{energiesrgb}
\end{figure}

To test whether the resolution affects our results, we have performed both
simulations with a higher resolution of $512^3$ cells.  The results are
qualitatively similar.  The main differences are that the orbital separation 
tends to a lower value in the higher resolution
simulation, and that the in-spiral is faster 
(see Fig.~\ref{multifigure}). 
In both cases, following the slow-down in orbital decay, 
the in-spiral continues at a slower pace until we stop the simulation. 
We also found that about half as much mass becomes unbound in the
higher resolution simulation ($\approx5\times10^{-4}~{\rm M_\odot}$ vs. 
$\approx9\times10^{-4}~{\rm M_\odot}$ for the lower resolution RGB simulation), 
despite the fact that at higher resolution more orbital energy is delivered
as the higher resolution allows us to follow the inspiral further.
The star remains somewhat more compact in the higher resolution simulation,
which is evident in the steeper increase in $V_{\rm p}$ as the planet
approaches the core, particularly in the RGB simulation.
The accretion radius of the planet 
is not resolved in any of our simulations. We discuss the implications
in Section~\ref{dragforcesubsection}.

Our simulations conserve energy reasonably well. We 
find that the total energy on the grid in the RGB simulation increases by 
$\approx3\times10^{46}~{\rm erg}$ over the course of the simulation (see
Fig.~\ref{energiesrgb}).  
This is $\approx3$ per cent of the initial gravitational potential energy of the
gas on the grid, which was $\approx1\times10^{48}~{\rm erg}$. It is $\sim$10
times the change in the planet's potential energy, and $\sim$100 times the
change in the planet's kinetic energy.  However, over the same time,
$8.8\times10^{-3}~{\rm M_\odot}$ are lost from the grid.  If all this lost
mass carried the thermal energy of the initial low density ambient medium,
this mass-loss from the domain would remove $6.6\times10^{46}~{\rm erg}$
from the grid\footnote{The kinetic energy and the gravitational potential
energy of the low density ambient medium are negligible compared to the
thermal energy.}.  We therefore estimate that the total energy has increased
by up to $9.6\times10^{46}~{\rm erg}$, corresponding to $10$ per cent of the
initial potential energy of the star.  We compare the energy gained due to
non-conservation to the potential energy of the star\footnote{The total energy of the star is almost identical to its potential energy, because kinetic and internal components are not very large.} instead of the total
energy in the box. The latter quantity is meaningless, because the total energy in the box can be made
arbitrarily high and close to zero by the addition of an arbitrary quantity of hot
``vacuum".

We also emphasize that 10 per cent is an upper limit to
the non-conservation, because most of the mass lost from the grid has a
lower thermal energy than the initial hot ambient medium. We find that much
of the gas leaving the simulation box has a specific thermal energy of
$\sim10^{13}~{\rm erg~g^{-1}}$. Assuming that this is representative for all
the gas leaving the box, we can determine a 
lower limit to the energy non-conservation. Then
$\approx2\times10^{44}~{\rm erg}$ would be lost from the box (i.e. a factor 
$\gtrsim100$ less than the above
estimate), and hence the energy non-conservation over the course of the
simulation would be approximately $3$ per cent. 
In assuming that the gas that leaves the box removes $2\times10^{44}~{\rm erg}$ we omitted
accounting for its kinetic and potential energies. However,
kinetic energy and gravitational potential energy of the stellar envelope
material lost from the grid have opposite signs and are of the same order 
of magnitude as the thermal energy. Hence, it will not significantly change
our estimate. Therefore, the
energy non-conservation in the RGB simulation is between 3 and 10 per cent,
likely closer to 3 per cent.
The conservation is slightly better in the higher resolution simulation, but
still has a lower limit of roughly 3 per cent. 

In the AGB simulation, we find that the total energy decreases by
$3\times10^{45}~{\rm erg}$ over
60 years.  However, the total potential energy in the AGB simulation is
$\approx8\times10^{46}~{\rm erg}$,
approximately a factor ten less than in the RGB simulation, since the AGB star is 
less tightly bound. 
The specific thermal energy of the ambient medium is similar to that in 
the RGB simulation, and therefore the level of non-conservation is even more
sensitive to how much thermal energy is carried away by the lost mass.
We find that $\sim1\times10^{31}~{\rm g}$ was lost from the grid over the course of the
simulation. If all of this mass had a specific thermal energy of
$\sim10^{13}~{\rm erg~g^{-1}}$, we find that the total energy should have
dropped $\sim1\times10^{44}~{\rm erg}$, which is small compared to the actual
drop of $3\times10^{45}~{\rm erg}$. This way, we find that the energy is
conserved to within $4$ per cent in the AGB simulation. This is, however, an
estimate for the energy conservation based on a
lower estimate for
the energy lost from the grid associated with mass loss. It is likely that
the lost mass has taken out a larger amount of thermal energy, which would
make the conservation better, unless the lost mass has removed more than
$\sim6\times10^{45}~{\rm erg}$ (corresponding to a specific thermal energy
of more than $\sim6\times10^{14}~{\rm erg~g^{-1}}$). We therefore expect the
energy to be conserved to within a few per cent also in this simulation.

\section{Drag forces}
\label{dragforcesubsection}

The torque acting on the planet dictates the rate of in-spiral. Determining whether simulations represent the drag forces with sufficient accuracy is an important step when determining whether results of simulations are reliable. 
Below we consider both gravitational and hydrodynamic drag components and compare what should be going on in nature, expressed by analytical approximations, with what is going on inside the simulation, calculated from the quantities that  are output from the code. 

\subsection{Gravitational vs. Hydrodynamic Drag}

In nature, a planet in a CE phase with its host star would experience a drag force composed of 
gravitational and hydrodynamic components. The hydrodynamic drag force is due to the ram
pressure on the planet from the surrounding gas, and this force can be
estimated:
\begin{equation}
F_{\rm hydro,drag}\sim \rho v_{\rm p}^2~\pi R_{\rm p}^2,
\label{hydrodrageq}
\end{equation}
where $v_{\rm p}$ is the planet's relative velocity with respect to the surrounding
gas, $R_{\rm p}$ is the radius of the  planet, and $\rho$ is the density of the envelope gas
surrounding the planet. 

The gravitational drag is instead due to gravitational forces between the gas flowing past the planet and the planet itself. Although there is no accurate expression for the gravitational drag in the presence of a density gradient \citep{MacLeod2015}, an approximate expression can be found in \citet{iben93}  and \citet{passy12}:
\begin{equation}
F_{\rm grav,drag}\sim \zeta \rho v_{\rm p}^2~\pi R_{\rm A}^2,
\label{gravdrageq}
\end{equation}
where $\zeta$ is a numerical factor that depends on the Mach number (it is larger 
than 2 for supersonic motion and less
than unity for subsonic motion \citep{shima85}), 
and $R_{\rm A}$ is the accretion radius given by \citep{iben93}:
\begin{equation}
R_{\rm A}=\frac{2GM_{\rm p}}{v_{\rm p}^2+c_s^2},
\label{rabondi}
\end{equation}
for subsonic and sonic
speeds, when pressure effects are included \citep{bondi52}. This tends to 
\begin{equation}
R_{\rm A}=\frac{2GM_{\rm p}}{v_{\rm p}^2},
\label{rahoyle}
\end{equation}
for high Mach numbers \citep{hoyle39}. In Eq.~\ref{rabondi}, $c_s$ is the 
sound speed. For simplicity we assume $\zeta=1$ always, which means that we
will underestimate the gravitational drag force in
the supersonic regime, and overestimate it in the subsonic one.

Our simulations do not reproduce the hydrodynamical drag, because the planet
is approximated by a point particle and has no surface.  Some hydrodynamic drag may be felt by the planet in the simulations due to the fact that some gas gathers in the potential well of the planet moving with it and in so doing it  collides with surrounding gas. However, because of the relatively low mass of the planet, this effect is small in the simulations.

Ricker and Taam (2008) predicted that the
hydrodynamic drag should be much weaker than the gravitational drag in common
envelope simulations with stellar-mass companions.  However, planets have
much lower mass and a correspondingly weaker gravitational drag.  As we show
in Fig.~\ref{multifigure}, towards the end of the RGB simulation the 
hydrodynamic drag should be comparable to or even dominate the gravitational
drag including pressure effects (which is the relevant gravitational drag
force at that time). At this point the simulations misrepresent the force on
the planet, and we stop them. This does not happen in the AGB
simulations, where the hydrodynamic drag should always be negligible compared to
the gravitational drag force.

\subsection{A Comparison Between Numerical and Analytical Expressions of the Gravitational Drag}

\begin{figure*}
\includegraphics[width=0.45\textwidth]{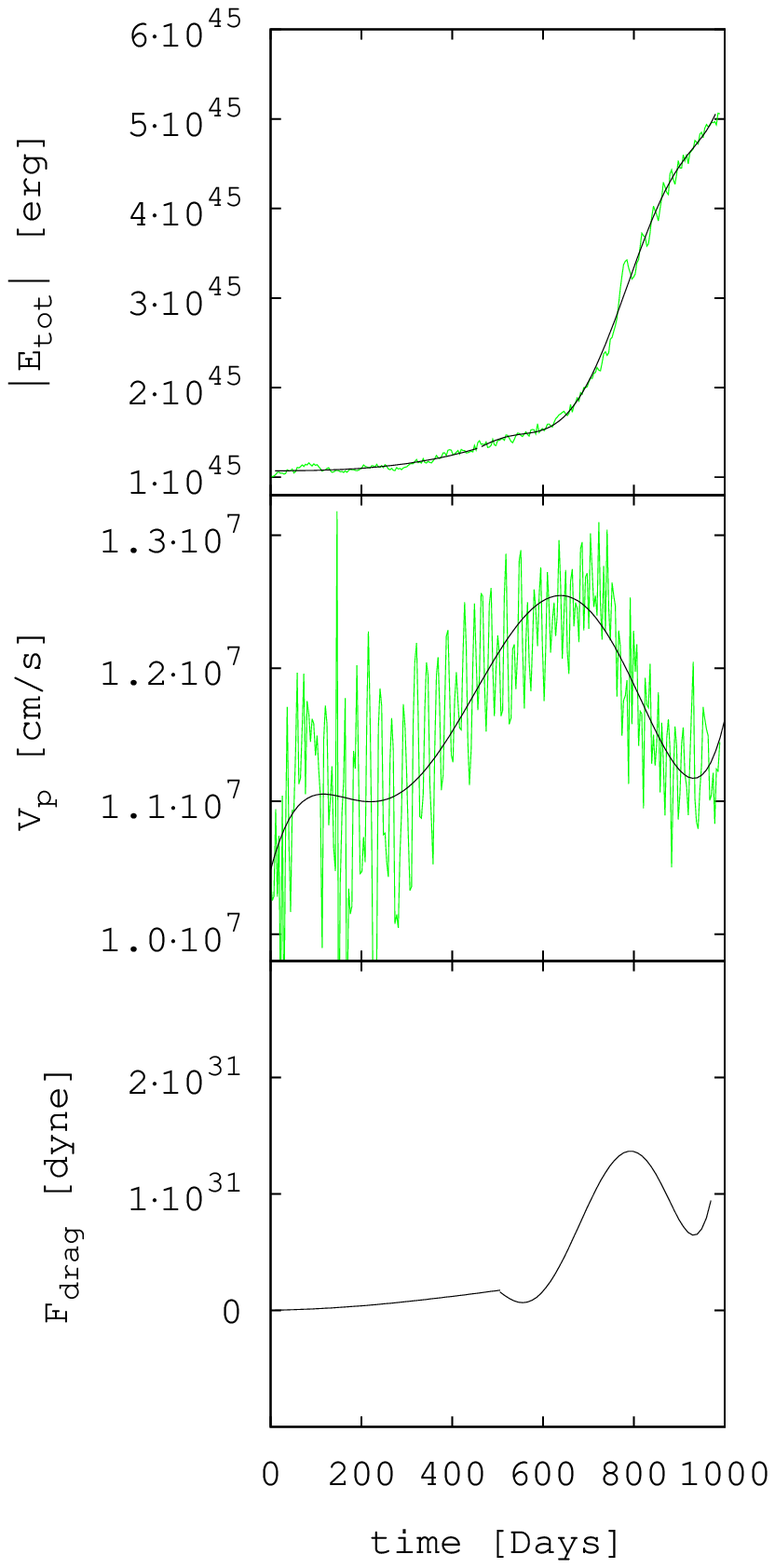}
\hspace{1cm}
\includegraphics[width=0.45\textwidth]{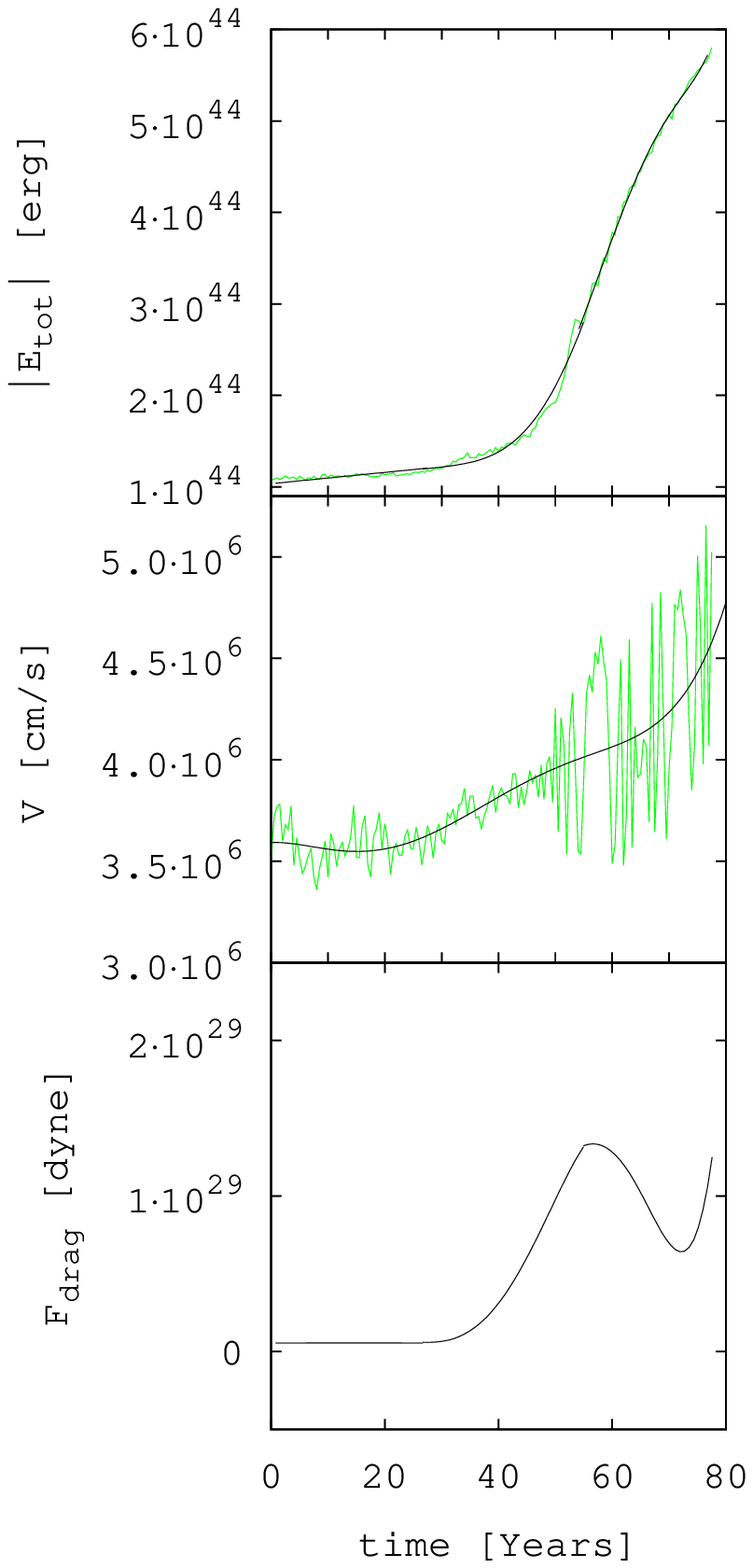}
\vspace{1.5cm}
\caption{Results from the low resolution RGB simulation (left column) and AGB simulation
(right column). The green curves are the raw data, showing {\it top panel:} 
the absolute value of the planet's total energy, {\it middle panel:} the
planet's velocity relative to the surrounding gas (in contrast to the
velocity relative to the grid shown in Fig.~\ref{multifigure}). The black 
curve is a fit to the total
energy (top panel) and the velocity (middle panel), and using these fits we 
have calculated the drag force (bottom panel). This drag
force is also shown in Fig.~\ref{multifigure}.}
\label{fitsfigure}
\end{figure*}

In order to compare the analytical estimates of the gravitational drag (Eqs. 2, 3 and 4) with the actual gravitational drag experienced by the planet in the simulations, we need to device a way to extract this information from the simulation outputs. We calculate the difference in the planet's energy (kinetic plus gravitational potential energies)
between two successive snapshots from the simulation. This difference is due to the gravitational drag force, which does work ($W$) on the planet. This force is approximately anti-parallel with the 
planet's motion, and its magnitude is therefore given by:
\begin{equation}
F_{\rm drag,code}=W/s,
\end{equation}
where $s$ is the distance travelled by the planet between two snapshots. We
estimate $s$ by taking the velocity of the planet at the first
snapshot and multiplying it by the time between the snapshots. 
The resulting drag force is plotted alongside the other relevant quantities in
Figs.~\ref{fitsfigure} . 

This estimate is approximate but reasonably accurate. We checked that this estimate of the force is similar to what would result from determining the orbit-averaged radial position of the planet at each time step, thereby determining the force by calculating the second differential of that radial distance and multiplying by the planet's mass. Another method is to read the total acceleration on the planet from the code output. This method is more noisy because the total value of the  acceleration includes the dominating centripetal value, which needs to be subtracted from the total.

The values of the orbital energy of the planet and of its velocity vary between one output frame and the next (see Fig.~\ref{fitsfigure}).  
The planet's total energy decreases but occasionally it grows
slightly between two snapshots, which results in
a drag force that is instantaneously negative.  In 
addition, the planet's velocity can vary by up to 
$25$ per cent between snapshots for the AGB simulation (see
Fig.~\ref{fitsfigure}, where we plotted the planet's velocity with respect to the
surrounding gas, in contrast to in Fig.~\ref{multifigure}, where we plotted the planet's
velocity relative to the grid).

To eliminate the oscillations we fitted the
total energy, as well as the
planet's velocity and use the fitted curves to determine the value of the gravitational drag force.  
The force curve has an upturn at the end of the curve, which is an edge effect inherited by the fit to the planet's velocity (middle panel in Fig.~\ref{fitsfigure}). In this figure, we only show
values for the lower resolution simulations.  More details about the
fits are provided in the appendix.

The gravitational drag force values
for both the lower and higher resolution simulations calculated with the method we have explained above, are also shown in the
bottom two panels of Fig.~\ref{multifigure} (black curve), where they are compared
to the hydrodynamic drag force (blue curve) from Eq.~\ref{hydrodrageq}, and to
the analytically-derived gravitational drag force (supersonic case: solid red curve, or including
pressure effects: dashed red curve).  We plot the gravitational drag force both
including and excluding pressure effects, because the planet is supersonic
until $\sim$800 days in the lower resolution RGB simulation and 
$\sim$55 years in the lower resolution AGB simulation (see the second panel in
Fig.~\ref{multifigure}). 

Both the expressions in Eqs.~\ref{hydrodrageq} and \ref{gravdrageq}, depend
on the density surrounding the planet and the velocity of the planet with
respect to the surrounding gas.  These are plotted in the 4th and 5th rows
of Fig.~\ref{multifigure}.  Initially, the planet starts just outside the
star at $1.1 R_{\rm star}$, and the density surrounding it is the low
ambient background density.  However, as the giant star expands, the planet
finds itself embedded in higher density material.  The few oscillations seen
in the density surrounding the planet at $\sim$200 days in the RGB
simulation and $\sim$20 years in the AGB simulation are due to the planet
acquiring a slight eccentricity, or because the star in our simulations is
not entirely spherical at this point in time, and so the planet may
encounter different densities even if it is in a circular orbit.  As the
planet in-spirals through the star's envelope, the density gradually
increases, to reach a maximum at the end of the simulation of $\approx
10^{-4}~{\rm g~cm^{-3}}$ after $\sim$1000 days, in the RGB simulation, and
$\approx 10^{-6}~{\rm g~cm^{-3}}$ after $\sim$80 years, in the AGB simulation.

The gravitational drag force calculated from the simulations follows closely
the pace of the in-spiral from which it is calculated.  It increases during
the fast in-spiral phase, between 600 and 800 days to $\sim
1-2\times10^{31}~{\rm dyne}$ in the RGB simulation, and between 30 and 50
years to $\sim 1-2\times10^{29}~{\rm dyne}$ in the AGB simulation.  This
leads to an acceleration of the planet due to the drag of approximately
$-1~{\rm cm~s^{-2}}$ in the RGB case, and $-0.01~{\rm cm~s^{-2}}$ in the AGB
case.  The difference in drag force in the RGB and the AGB simulations is
primarily due to the different densities encountered by the planet.  The
peak force in the higher resolution simulations is approximately a factor of
two larger than in the lower resolution simulations (see
Fig.~\ref{multifigure}).

Looking at the drag forces in the last panel of Fig.~\ref{multifigure} -- for
the high resolution simulations, we see that the computationally-derived
force is 2-3 times larger than the analytically-calculated {\it
supersonic} gravitational drag force for both RGB (at around 600 days) and AGB
(between 40 and 50 years) simulations, due
either to the use of $\zeta = 1$ in Eq.~4, or to an actual effect of the
density gradient exerting an added component to the force
\citep{MacLeod2015}.  Following this increase, the drag force then decreases
becoming the same as the expression for the gravitational drag including
pressure effects at 820 days and 65 years for the RGB and AGB simulations,
respectively.  The force values peak when the planet's Mach numbers are
slightly larger than unity in both simulations.  We interpret this by
looking at Eq.~\ref{gravdrageq} and noticing that in the supersonic regime,
$R_{\rm A}$, and the gravitational drag force, can be seen to increase with
decreasing velocity (smaller Mach numbers).  However, when the force
transitions to the sonic case, $\zeta$ is smaller, and hence the force
decreases.  Although a quantitative comparison cannot be carried out, this
behaviour is that predicted by \citet{Ostriker1999}:  the force is 
largest at, or slightly above the sonic point and drops dramatically just 
below Mach-1. In these simulations we conclude this is what causes the 
sudden decrease in drag force
that makes the in-spiral slow down (see Fig.~\ref{multifigure}).  

We finally note that the particle representing the core of the giant also experiences a drag force. However, this particle moves very slowly, much slower than the local sound speed. As the particle is very massive, it affects its surroundings significantly, for instance by attracting mass. It may therefore be difficult to accurately determine the relative velocity between the particle and its surroundings. We have not made any attempts to calculate the force acting on it.

\subsection{The dependence of simulated gravitational drag on resolution}

Presumably the difference between lower and higher resolution simulations has something to do with the strength of the interaction which takes place in the vicinity of the planet. It is likely that the strength of the interaction is not well represented, for example, when $R_{\rm A}$ is not well
resolved. In our higher resolution
simulations $R_{\rm A}$ is a factor 3-8 smaller than the cell
size, while in the lower resolution simulations it is a factor 5-15 smaller
than the cell size (the range is due to the fact that $R_{\rm A}$
varies, while the cell size is fixed). In principle a convergence
test carried out over multiple resolutions with a range of smoothing lengths
could identify a problem, but due to the computational
expense of these simulations, only limited resolution tests are carried out. 
When limited tests are carried out several effects can counter each other
and confuse the issue of whether the gravitational drag is well represented.

We have performed several test simulations of a point mass moving through a
constant density medium in order to eliminate the density gradient, which
can complicate matters, and to reduce the computational expense of the tests. 
In these test simulations we have varied the density, thermal energy
(and hence the sound speed), and the particle's velocity.  In this way we
have been varying $R_{\rm A}$ so as to make it larger or smaller than the
cell size.  We found that when $R_{\rm A}$ is under-resolved, the drag force
acting on the particle in the simulation tends to be overestimated compared to the
analytical expression.  

Contrary to this expectation, we find that the
peak drag force is a factor $\approx2$ {\it higher} in the high resolution
simulations than in the low resolution simulations.  It is possible that this may be due to 
the density structure of the 1D {\sc mesa} model in the 3D grid being 
more compact and have higher density in the higher resolution (Fig.~\ref{multifigure} 
fifth panel from the top), compared to the lower resolution simulations.  
The more compact and dense giant star in the higher resolution simulation means that
not only the peak drag force, but the drag force acting on the planet in general 
is larger than at lower resolution.
This may make up for the possible force under-estimation due to not resolving $R_A$.

\section{Discussion}
\label{summarysection}

\subsection{The intensity, timescale and frequency of planet merger transients}

We start by examining the interaction timescales. We find that the planet
in-spirals relatively fast (few years for RGB stars and $\sim$100 years for
the AGB case), although this is slow compared to CE interactions with more
massive companions \citep[e.g.,][]{ricker12,passy12,DeMarco2012}.  The hydrodynamic drag was
not modelled, but we found that it could play a role at the end of the RGB
interaction and this could shorten the in-spiral timescale somewhat. 
Pre-empting our discussion in Section~\ref{ssec:destroying} the planet will
be eventually destroyed on a timescale that is likely of the same order of
magnitude as the one characterising the initial in-spiral.

During the time of the in-spiral, the photosphere expands and the star likely brightens. 
Using the values of the density at the photosphere from the \textsc{mesa} 
models ($\approx9\times10^{-9}~{\rm g~cm^{-3}}$ and  
$1.6\times10^{-9}~{\rm g~cm^{-3}}$ for the RGB and AGB stars, respectively) 
we find that the interaction with the planet has caused the RGB star to 
expand by $\approx 40$ per cent over 3 years, and the AGB star 
to expand by $\approx 20$ per cent over 80 years. If the 
temperature of the photosphere remains
constant, this would indicate a modest increase in luminosity by a factor of two for
the RGB star, and $\approx40$ per cent for the AGB star. The temperature however, may decreases somewhat, as is demonstrated by Mira stars that can double
their radius and halve their effective temperature over pulsation cycles of
a few $\times$100 days (e.g., $o$~Cet; \citealt{Ireland2008,Ireland2011}). Additional cooling of the photosphere may be expected in the case of the AGB star. If we accounted for a decreasing temperature linearly inverse to the increase in radius then the luminosity would actually drop.
It is possible, on the other hand, that the photosphere would be farther out than we have considered because of the low density
material that readily expands out. Unfortunately we cannot integrate the optical depth of the material because  its temperature is affected by the artificially large ``vacuum" temperature used in {\sc enzo}, making these tenuous outer layers more optically thick than they should be. 

The average thermal timescales of the stars are 7600 and 30 years for the
RGB and AGB stars, respectively.  The RGB simulation ends at $\approx$4~\rsun,
and this is likely an upper limit.  It is entirely possible that with a
higher resolution, the ``destruction depth" of $\sim 1$~\rsun\
(Sec.~\ref{ssec:destroying}) would be reached within similar
timescales.  For the AGB star this is less likely.  However its thermal
timescale is much shorter and of the order of the in-spiral timescale.  It
is therefore possible that the AGB star would contract on the same timescales as
it is expanding because of the injected orbital energy.  If this happened,
it is likely that the in-spiral would continue.  We posit therefore that
both interactions would result in the planet destruction within a timescale
that is of the same order of magnitude of the in-spiral timescale.

Assuming that there are $10^{10}$ stars in the Galaxy that are able to evolve 
off the main sequence over the age of the
universe, and that they have an average lifetime of 10 billion years, then we would have 
$10^7$ RGB and $10^6$ AGB stars at any given time in the Galaxy (using RGB and AGB lifetimes of 10 and 1
million years, respectively - see \citealt{Moe2006} for references to this
back of the envelope calculation).  Given the planet-swallowing timescales
determined in this work, this would mean that one RGB star in a million
would be undergoing an interaction with a companion if all RGB star went
through one such interaction in their lives.  For the AGB it would be one
star in 10\,000 if they too went through one such interaction in their
lives.  This would mean that 10 RGB stars and 100 AGB stars in the Galaxy
would be going through such interaction at any one time. These predictions are similar to what could be surmised by the considerations of
\citet{metzger12} who discuss that the rate of planet-main sequence star merger should be similar to the rate of planet-giant star merger, both approximately a few per year.

Given the long brightening and dimming timescales (relatively to any survey timescale) and
the relatively small variation amplitude predicted, these phenomena may not
be observable, unless a more powerful outburst could be triggered
\citep{soker91,Bear2011}. This would be quite different from the case of a planet-main sequence star merger discussed by
\citet{metzger12}.  Alternatively, as discussed in \citet{nordhaus06}, if the planet is
tidally disrupted, it can form a disc deep inside the star which can lead to
a disc driven outflow.

Very little mass is unbound from the system.
Energy loss due to non-conservation may have decreased the mass-loss rate 
from the AGB star somewhat, but this could not be said of the  RGB star for 
which the total energy slightly increased due to lack of perfect conservation. 
Additionally, non-simulated effects that could increase the mass-loss rate 
may be the interference, particularly for the case of the AGB star, of the 
orbital period with the fundamental pulsation period of the star. If little 
or no mass is ejected from the system due to the interaction, it is likely 
that the stars will settle back into an equilibrium stage after radiating 
their excess energies over their thermal timescales of
7600 and 30 years, for the RGB and AGB stars, respectively.

The interaction caused the puffed-up, low density, outer layers of the star
to rotate, with velocities $>20~{\rm km~s^{-1}}$ in the RGB star and
$>5~{\rm km~s^{-1}}$ in the AGB star (the extra angular momentum transported
by a planet captured tidally would only change these values slightly).  At
higher densities, we found no significant rotation.   This could
indicate that the differential rotation mediated dynamo effect suggested in
\citet{nordhaus06} will not lead to large scale outflows. We expect that as the
interaction ends and the star settles back into its original configuration,
and the angular momentum is redistributed in the star, the surface rotation
would slow down.  Hence, an apparently relatively fast spinning giant star
for a brief period could be an indication of a recent CE interaction between
the star and a giant planet.

\citet{carlberg09} investigated the ability of planet accretion to
spin up stars, and found that in some cases RGB stars could become rapid
rotators due to merger with a companion planet although they found that fast rotation was more
likely to be achieved if the planet was captured by a sub-giant, as stronger
mass loss from giants can remove angular momentum from the envelope
preventing the rapid rotation. If this happened, the giant would be slowly
rotating or not rotating at all. Based on our simulations we suggest that the CE
event is still capable of causing a rapid rotation in the outer puffed up
envelope, as the CE interaction is fast and mass loss therefore can not remove
angular momentum sufficiently fast to prevent the spin-up.

\subsection{Destroying Planets and Polluting Giant Stars}
\label{ssec:destroying}

\begin{table}
\begin{tabular}{lcc}
\hline
 & RGB star & AGB star\\
 \hline
 Mass (\msun) &3.5 & 3.0\\
 Envelope mass (\msun) & 3.0 & 2.5\\
 Hydrogen mass (\msun) & 2.1 & 1.75\\
Enrichment@[Fe/H]$_\odot$ (\%) & 0.8 & 1\\
Enrichment@[Fe/H]=-1.7 (\%) & 43 & 50\\
\hline
\end{tabular}
\caption{Increase in the mass fraction of iron assuming that the destroyed planet has a core made of iron with a mass of 10~M$_\oplus$.}
\label{table1}
\end{table}
We assume that at some point the planet will be destroyed in the envelope of the giant star.
During the in-spiral there are competing processes that try to disrupt
the planet.  These act on different timescales and vary with
depth.  The planet can be (i) disrupted by shear between its outer layers and
the stellar ambient density, (ii) it can he ablated by heating and (iii) it can be
tidally disrupted.  We find that the planet is stable against
Kelvin-Helmholtz and Rayleigh-Taylor instabilities caused by shear
\citep[discussed in][]{passy12planet} for the conditions prevailing during our simulations. 
We find that a $10~{\rm
M_J}$ planet will be ablated by heating when the separation between the
planet and the core of the giant star is $\sim1~{\rm R_\odot}$ \citep{soker98}.  This is
also the distance from the stellar core at which the
planet will overflow its Roche lobe.  This is a much smaller separation than
the values of 10 and 85~\rsun\ reached at the end of our RGB and AGB
simulations, respectively.  Therefore we presume that this event has not yet
taken place, but will in time (a time possibly commensurate with the time for the early in-spiral).

Next we ask whether massive planets such as those we have simulated, once
destroyed at $\sim$1~\rsun\ can alter the giant composition in an observable
way.  The masses and compositions of the cores of massive exoplanets are
poorly known \citep[especially for hot Jupiters; for a recent review,
see][]{spiegel14}.  We assume that a $10~{\rm M_J}$ planet consists mainly
of an atmosphere of hydrogen and helium in solar proportions, and of a core
with an iron mass of $m_{\rm Fe}=10~{\rm M_\oplus}=3\times10^{-5}~{\rm
M_\odot}$ \citep{guillot99}.  
The base of the convective region in our \textsc{mesa} RGB model
is at $\approx0.4~{\rm R_\odot}$ while for the AGB star it is at $\approx0.2~{\rm
R_\odot}$; both are deeper than the location at which we predicted the planet to be destroyed. The disrupted planet mass will
therefore quickly be mixed into the giant stars' envelopes due to convection.

In Table~\ref{table1} we list the RGB and AGB star masses, envelope masses, and hydrogen masses for a hydrogen mass fraction of 70 per cent.
For a Solar metallicity
\citep[$\epsilon_{\rm Fe}=7.47$ for the Sun or $m_{\rm Fe}/m_{\rm H}\approx 0.0017$;][]{scott15} we therefore see that the
added iron from the planet increases the envelope metallicity too little to be observed.

If we assumed that the iron mass fraction has to grow by at least a factor
of 1.5 to be discerned from the base metallicity of the star, then the base
metallicity of the star should be [Fe/H]$<-1.7$ in the AGB case.
A giant with a mass of $\sim$1~\msun\ and an
envelope mass of 0.5~\msun\ would enable us to detect the pollution
more readily at higher, but still sub-solar metallicities ([Fe/H]$<-1.3$).  

Since there appears to be a correlation between a planet's metal 
fraction (i.e., core mass in a gas giant) and the
metallicity of the host star \citep{guillot06}, it may be that
such low metallicity stars cannot harbour metal rich planets. 
On the other hand, there may also be considerable variability in the metal content of planets.  For
instance, the planet HD 149026b is thought to contain $60-93~{\rm M_\oplus}$
of heavy elements \citep{fortney06}, much more than we have considered above.  
However, even such large core mass
would not be able to noticeably alter the observed metallicity of a
Solar metallicity star.

Another possibility for getting metal enrichment in AGB stars was discussed
by \citet{soker92}, who studied common envelope interactions between AGB
stars and brown dwarfs, and suggested that for separations between $3-10~{\rm
R_\odot}$, the brown dwarf would excite gravity waves that could lead to a
spin up of the inner envelope.  This could also lead to mixing near the
core, causing extra dredge-up of core material into the envelope.  Hence, if this
process happened near the last stages of mass loss, the wind of the AGB star
would be enriched in heavier elements.  However, this star would be a much more evolved
AGB star than the one we have considered in this work, and this mechanism requires that the companion enters the common envelope only at the very late stages of AGB evolution.

\section{Summary}

We have simulated the CE interaction between a $10~{\rm M_J}$ planet and a
3.5~\msun\ RGB star or a 3.05~\msun\ AGB star using the grid code Enzo with a
uniform, cubic grid with a maximum resolution of 512 cells on a side.
These
simulations have several limitations, but can give order-of-magnitude
quantitative information.

The limited resolution in our simulation affects
the final separation of our simulations, and some of the results from late
times in our simulations may not be accurate. Another effect of the resolution is
that the accretion radius is not resolved, which can lead to an overestimate
of the force. However, in lower resolution simulations the star diffuses out
more leading to lower densities which can cause an underestimate of the
force, somewhat counteracting the overestimate from not resolving the
accretion radius. Future simulations using an adaptive mesh refinement
simulation code, or possibly a smoothed particle hydrodynamics code may be
able to overcome some of these limitations. We nevertheless found that:

\begin{itemize}

\item Plunge-in times of the order of years to decades are seen in our 
simulations for the RGB and AGB cases, respectively. The plunge-in times of low mass 
companions such as planets in the envelopes of giants are relatively longer 
than for more massive, stellar companions, with the longer times being 
witnessed for the more evolved, lower density primaries. 

\item We concluded that the planets should not be disrupted during the simulated phase. We cannot tell with precision how much longer the planets will take to reach a depth where disruption takes place.

\item Destroyed planets will pollute the envelopes of giant stars, but the 
effect is likely to be witnessed only in the lowest mass giants with the 
lowest metallicity, if these stars can have planets with suitably massive metal cores.

\item Only a very small amount of the primary star's envelope mass is
unbound by the planet in our simulation.  It is possible that if the planet
interacts with the star's pulsation this may trigger further unbinding, or,
if the planet is tidally disrupted it can form a disc inside the giant star
from which a disc driven outflow can form.
 
\item The expanding giant's luminosity may increase by a modest factor over
a relatively short timescale of the early in-spiral (though still long
compared to survey timescales).  This effect would likely be relatively rare
and difficult to observe.

\item In line with other studies we find that the penetration of the planets into the giants will stimulate faster rotation. However as this rotation is limited to the outer layers, it is not clear in what timescales the angular momentum will re-distribute into the entire envelope and what the final rotation rate of the giants will be. 

\item Analytically, it is predicted that the gravitational drag force would
 peak at the sonic point and
greatly diminish for sub-sonic regimes.  In our simulations the slowing down of the
in-spiral takes place at such a transition.  The overall force experienced
by the planets in our simulations is larger than calculated analytically and
is larger for higher resolution. This may simply be due to us assuming
$\zeta=1$ in Eq.~\ref{gravdrageq}.  It is also possible that the presence of a
density gradient may enhance the intensity of the gravitational drag.  We
leave further comparisons between numerical and analytical gravitational
drag to future work.

\end{itemize}

\section*{Acknowledgements}
We thank the anonymous referee for constructive comments which helped
improve the paper.
We thank B. Pandey and M. Wardle for constructive discussions during this
work, and C. O'Neill for helpful input.
J.E.S acknowledges support from the Australian Research Council Discovery 
Project (DP12013337) program.
O.D. gratefully acknowledges support from the Australian Research Council
Future Fellowship grant FT120100452.
J.-C.P. acknowledges funding from the Alexander-von-Humboldt Foundation. 
This research was undertaken, in part, on the NCI National Facility in
Canberra,
Australia, which is supported by the Australian Commonwealth Government.
Computations described in this work were performed
using the Enzo code (http://enzo- project.org), which is the product of a
collaborative effort of scientists at many universities and national
laboratories.

\begin{appendix}

\section{Fitting results}

For both of the simulations, we fit the planet's velocity and the planet's
total energy, to get smooth curves.  We have not attempted to estimate the
$\chi^2$ of the fits.  Instead, we have shown in Fig.~\ref{fitsfigure} the
data and the fitted curves from the low resolution simulations, and limit
ourself to stating that qualitatively the fits look reasonable.  In this
appendix we show the details of the fits and the results for the two low
resolution simulations.

\subsection{RGB case}

We fit the planet's velocity with a fifth order polynomial
($ax^5+bx^4+cx^3+dx^2+ex+f$) over the entire data-range from 0 to 980 days.
The result of the fit is: $a=9.85028\times10^{-8}$, $b=-0.000237$,
$c=0.18886$, $d=-56.0575$, $e=7229.48$, and $f=1.04192\times10^7$. The
planet's negative total energy was fit with two different curves. To ensure a reasonably continous fit
with a reasonably continuous first derivative, we fit the curves over a
larger range than we plot them. From 0 to 560 days
we use a third order polynomial ($f_1(x)=a_1x^3+b_1x^2+c_1x+d_1$)
where we found the coefficients to be $a_1=1.93253\times10^{36}$,
$b_1=3.2207\times10^{38}$, $c_1=2.28898\times10^{39}$, and
$d_1=1.07075\times10^{45}$. This was used to plot the curve from 0 to 510
days. Then from 400 days to 980 days we used a sixth order polynomial
($f_2(x)=a_2x^6+b_2x^5+c_2x^4+d_2x^3+e_2x^2+f_2x+g_2$), where we found the
coefficients to be $a_2=4.21665\times10^{30}$, $b_2=-1.73393\times10^{34}$,
$c_2=2.89879\times10^{37}$, $d_2=-2.51962\times10^{40}$,
$e_2=1.20186\times10^{43}$, $f_2=-2.98564\times10^{45}$, and
$g_2=3.03351\times10^{47}$. This was plot from 510 days to 980 days.

\subsection{AGB case}

We fit the planet's velocity to a fifth order polynomial
($ax^5+bx^4+cx^3+dx^2+ex+f$) over the entire data-range from 0 to 77.5
years.  The result of the fit is: $a=0.00832347$, $b=-1.52515$, $c=96.5916$,
$d=-2287.46$, $e=22336.7$, and $f=3.35088\times10^6$, with 150 degrees of
freedom.  The planet's negative total energy we fit with three different
curves.  To ensure a reasonably continuous fit with a reasonably continuous
first derivative, we fit the curves over a larger range than we plot them. 
From 0 to 35 years we fit a first order polynomial ($f_1(x)=ax+b$)
 and results in $a=5.28612\times10^{41}$ and $b=1.05581\times10^{44}$.  This
we used to plot from 0 to 27 years.  From 5 to 70 years we fit a sixth order
polynomial ($f_2(x)=ax^6+bx^5+cx^4+dx^3+ex^2+fx+g$) which results in
$a=-1.22317\times10^{35}$, $b=2.12089\times10^{37}$,
$c=-1.30396\times10^{39}$, $d=3.64203\times10^{40}$,
$e=-4.51527\times10^{41}$, $f=2.24252\times10^{42}$, and
$g=1.07388\times10^{44}$, which was plotted from 27 years to 55.5 years. 
Finally, from 45 to 77.5 years we fit an eight order polynomial
($f_3(x)=ax^8+bx^7+cx^6+dx^5+ex^4+fx^3+gx^2+hx+i$), which was fitted from 45
to 77.5 years and results in: $a=3.45742\times10^{31}$,
$b=-8.18603\times10^{33}$, $c=7.09231\times10^{31}$,
$d=-2.71983\times10^{37}$, $e=4.27398\times10^{38}$,
$f=5.97299\times10^{22}$, $g=1.12244\times10^{21}$,
$h=2.26162\times10^{19}$, $i=1.24344\times10^{16}$.  This was plotted from
55.5 years to 77.5 years.

\end{appendix}

\end{document}